# THREE DIMENSIONAL FILAMENTATION ANALYSIS OF SDSS DR5 SURVEY

**Yongfeng Wu**
Department of Astronomy, University of Science and Technology of China,
Jinzhai Road 96, Hefei 230026, Anhui Province, P.R.China
**David J. Batuski**
Department of Physics, University of Maine, Orono, ME 04469, USA
**Andre Khalil**
Department of Mathematics & Statistics and Institute for Molecular Biophysics,
University of Maine, Orono, ME 04469, USA
yongfeng.wu@maine.edu

## Abstract

We introduce a new method to calculate the multi-scale 3D filamentation of SDSS DR5 galaxy clusters and also applied it to N-body simulations. We compared the filamentation of the observed vs. mock samples in metric space on scales from 8 Mpc to 30 Mpc. Mock samples are closer to the observed sample than random samples, and one of the mock samples behaves better than another one. We also find that the observed sample has a large filamentation value at a scale of 10 Mpc, which is not found from either mock samples or random samples.

## 1. Introduction

From redshift surveys such as the Sloan Digital Sky Survey (SDSS; York et al. 2000) and the Two-Micron All Sky Survey (2MASS; Skrutskie et al. 2000), the local (few to many tens of Megaparsecs) Universe shows intricate patterns with clusters, filaments, bubbles, sheet-like structures and so-called voids. For a review of the structural analysis of the Universe, see Weinberg (2005). At the same time, Lambda Cold Dark Matter (LCDM) models have been developed; see Gill et al. (2004) and Dolag et al. (2008). Several simulations incorporating dark energy have been created, such as the Millennium Simulation done by Croton et al. (2006) and another N-body simulation by Berlind et al. (2006). These models describe a Universe that consists mainly of dark energy and dark matter and calculate the evolution of the Universe from a short time after the big bang to the present time. As complicated evolution systems are sensitive to the initial conditions (Chen et al, 2011, Wang et al, 2011), the initial conditions of those simulations are strictly limited by current observations. Work has been done to verify the similarity between the real Universe and the simulated Universe (Springel et al. 2005; Berlind et al. 2006; Wu, Batuski & Khalil 2009) and they correspond well, based on the comparative techniques used in these studies.

To supplement the widely used correlation function and power spectrum (Yang et al, 2001, Cao et al, 2006), alternatives have been proposed to quantify structure in the galaxy distribution, such as the genus curve (Zeldovich 1982), percolation statistics (Zeldovich 1982; Shandarin 1983; Sahni et al. 1997), Rhombic Cell analysis (Kiang, Wu & Zhu 2004), void probability functions (White 1979), high-order correlation function (Peebles 1980), and multi-fractal measures (Saar et al. 2007). Filamentation is a traditional way to describe the structure of the galaxy distribution and measures of this property are widely used in the research of the real universe and simulations (Somnath et al. 2000). In this paper, we consider a wide range of smoothing levels for multi-scale filtering (Wu, Batuski & Khalil 2009). By varying the size of the smoothing function over a range of scales, a complete multi-scale filament form description of galaxy distributions becomes possible. Key facets of our filamentation approach are consideration of any given map as an element in the space of all such maps and definition of a distance function in that space to make the space of all maps into a topological space (Adams 1992). Moreover, the other methods just listed focus on summary statistics that convey little of the geometric and topological properties of the galaxy distribution. Our method also gives desired quantitative summary statistics of the difference between maps. However, a primary benefit of our method is that the filament function is straightforward and simple to understand and particularly useful in map comparisons.

## 2. 3-D filamentation analysis

### 2.1 Filament Function Definition

First we summarize the 2-D filamentation approach (Wu, Batuski & Khalil 2009), the definition of the diameter $D$ of a set $G$ is:

$$D(G) \equiv \max\{|x-y|, x, y \in G\} \quad (1)$$

Components are defined as isolated high-density regions in the map. The size shape and number of components will vary as a function of threshold values (Wu et al, 2009).

The filament index previously used in our 2D analysis is defined as:

$$F = \frac{PD}{4A} \qquad (2)$$

where $P$ is the perimeter, $A$ is the area and $D$ is the diameter. Now we define the 3-D filament index

$$F = \frac{SD}{6V} \qquad (3)$$

where S is the component's surface, V is the volume & D is the diameter.

This definition of the filament index satisfies intuitive requirements:

(1) The index should be proportional to D.

(2) The index should be inversely proportional to volume, with fixed surface and diameter. The fatter the object is the smaller index it should have. In other words, we can increase the volume and maintain the diameter and surface values (the surface increased on the body has been cancelled out by the surface decreased by the reduced spikes) (see Figure 1).

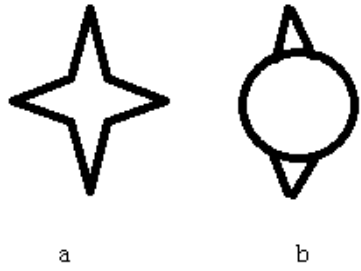

Figure 1. The filament index of a is greater than the filament index value of b in these 2-D views of 3-D objects.

(3) The filament value should be proportional to the surface. With fixed diameter and volume, the larger the surface is, the larger filament value it should have, as in Figure 2.

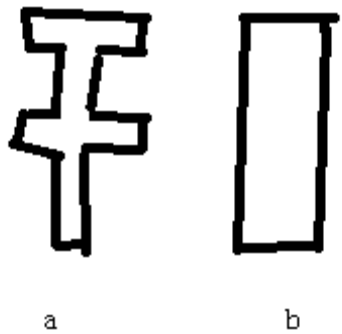

Figure 2. The filament index of a is greater than the filament index of b in 2-D views of 3-D objects.

Therefore the filament index can be used to quantitatively characterize the complexity of the object.

### 2.2. The distance between maps and Multi threshold values

If we want to compare filamentation between two maps, we define their metric distance as:

$$d_k(\sigma_A, \sigma_B) = (\sum | K(\sigma_A; \Sigma) - K(\sigma_B; \Sigma) |^p)^{1/p} \qquad (4)$$

where Σ is the threshold value, from minimum to maximum voxel intensity (Robitaille et al 2010), and p=2. We only keep pixels above the threshold value once a threshold value is defined for a map. $K$ is the filament function and $\sigma_A$ and $\sigma_B$ are maps. Multi threshold values can give us a full understanding of distance of two maps. However, different threshold values set can possibly get different distances. Here we use 10 threshold values equal spaced from maximum to minim value of the map. The reason is because we think (1) 10 threshold values are enough to fully describe the map (2) there is no reason to give some specific thresholds different weight than others.

In order to obtain the distance between the filament functions of the images under study, in this paper we apply this method in two ways. One way is that the observed images are compared to uniform images; giving us information on "how far" the samples fall from uniformity, thus giving quantitative information on the complexity of the observed images. Another way is that all simulation images are compared to SDSS observed images, thus, each measured distance gives quantitative information as to "how far" the simulation image is from observed data sets. Clearly, the larger the distance is, the "farther" the simulation image under study is from the observational data. The distances are calculated for the filamentation function, for each of the mock sample data sets, and for each size scale considered.

## 2.3. Gaussian smoothing and Multi scale analysis

The 2-D Gaussian smoothing function (equation (5)) is:

$$G(x, y) = \exp(-|\mathbf{x}|^2 / 2) \quad (5)$$

where $|\mathbf{x}| = \sqrt{x^2 + y^2}$ is a smoothing length, it governs the level of smoothing of the discrete data. The smoothing length obviously influences the structure analysis: underestimated smoothing length will cause huge numbers of false oscillations, but overestimated smoothing length will remove real features of structure. Figure 3 is an example of Gaussian smoothing.

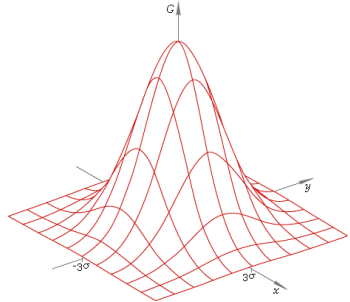

Figure 3. Gaussian smoothing function. (R. Fisher, 2003)

Gaussian filtering can be described by

$$T_G |f|(\mathbf{b}, a) = \frac{1}{a^2} \int f(x) \cdot G(\frac{\mathbf{x} - \mathbf{b}}{a}) d^2 x \quad (6)$$

where $f$ is a two-dimensional function representing the image under study, $G(\mathbf{x})$ is the Gaussian function (Equation 4), which can also be defined as a wavelet. $a$ is the scale parameter, and $\boldsymbol{b}$ is a position vector. Thus, the convolution between the point distribution images under study and the Gaussian filter at several different values of the scale parameter *a* yields the continuous gray-scale images from which the output functions and then the metric space coordinates can be calculated.

Gaussian filtering results in images with different filtering scales. In this paper we use a set of smoothing lengths from 10 Mpc to hundreds of Mpc. Figure 4 is a 2-D example of sketching this process.

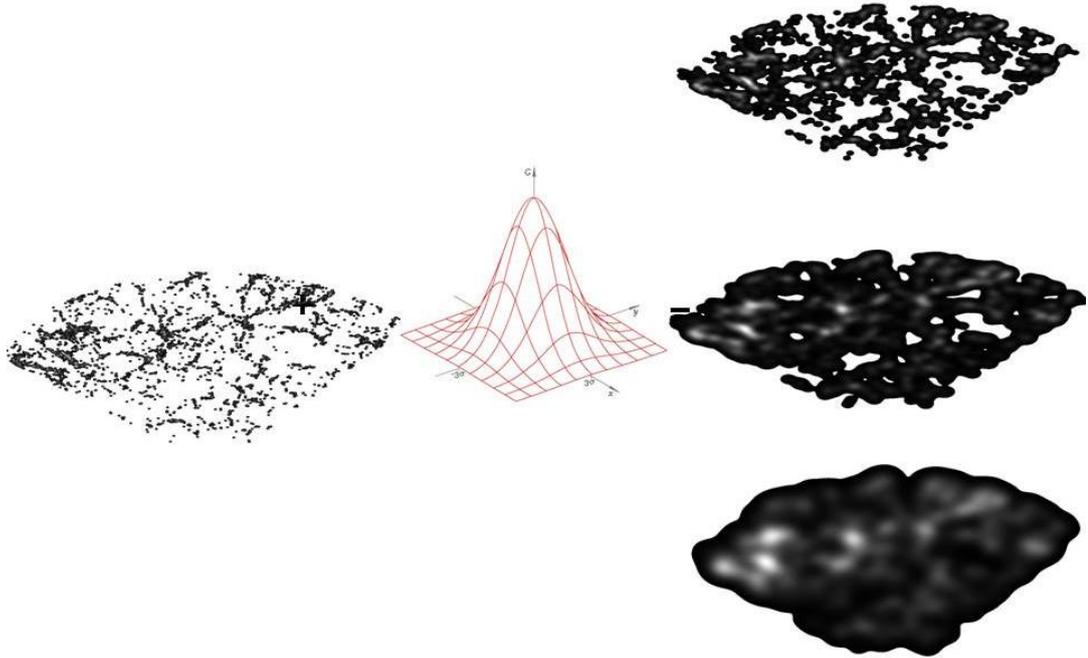

Figure 4. Gaussian filtering sketch map, the right three figures have increased smoothing length from top to bottom, we can see they have larger and larger clumps.

Multi-scale analysis is then possible with the using of different Gaussian smoothing length. We can extract specific scale components after smoothing with specific length. Multi-scale analysis is important in the geometry analysis of galaxy distribution as the geometry property is general different on different scales.

## 3. Data

**3.1 Observed data**

We use the SDSS Data Release 5 as our galaxy sample. We restrict our sample to regions of the sky where the completeness (ratio of obtained redshifts to spectroscopic targets) is greater than 90%, redshift range is 0.015 - 0.1 and $-48.3° < \lambda < 48.5°$ and $6.25° < \eta < 36.25°$ ($\lambda$ and $\eta$ are the telescope coordinates). Our final sample covers 2904 $\deg^2$ on the sky and contains 406594 galaxies (~40,000 galaxies after applying volume-limiting selection, as in the next paragraph).

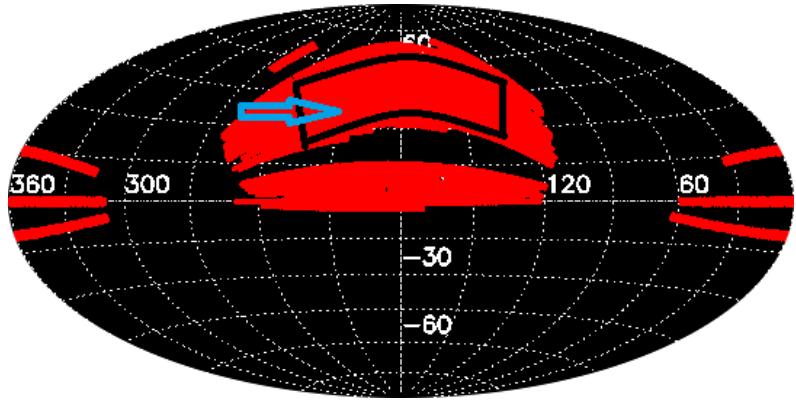

Figure 5. SDSS sample geometry. The region inside the black "rectangle" of the right figure is what we used.

We use volume limited (VL) samples (e.g., Davis & Peebles 1983), by choosing an upper cutoff in distance and calculating the absolute magnitude *M* according to the apparent magnitude limit of the telescope and this upper cutoff. The relationship between a galaxy's apparent magnitude and absolute magnitude is given by the expression

$$M = m - 5\log d + 5 \qquad (7)$$

*M* is the absolute magnitude, *m* is the apparent magnitude, and *d* is the distance from the observer. We only keep those galaxies whose absolute magnitude value is smaller than (brighter than) *M* for our faintest detectable galaxy at our redshift limit, this will ensure the selected galaxy sample is substantially complete to our magnitude limit.

**3.2 Redshift-distance formula**

From Weinberg (Weinberg 1972, Page 42, we neglect $\Omega_R$ (radiation) in the current matter-dominant Universe):

$$d_L = H_0^{-1}(1+z)\frac{1}{\Omega_k^{1/2}} \sin n\{|\Omega_k|^{1/2} \times \int_0^{z_1}[(1+z)^2(1+z\Omega_M) - z(2+z)\Omega_\lambda]^{-1/2}dz \quad (8)$$

Here $\Omega_k = 1 - \Omega_M - \Omega_\lambda$, $H_0$ is the Hubble constant, $z$ is the redshift, $z_1$ is the object redshift, $d_L$ is the luminosity distance (distance based on luminosity or magnitude). The *sin n* function is *sinh* function when $\Omega_k > 0$ (open Universe). It is only *sin n* when $\Omega_k < 0$ (closed Universe). When $\Omega_k = 0$, all terms include $\Omega_k$ will disappear. Equation (6) is used to calculate the distance of SDSS samples.

**3.3 Mock samples**

Our first mock sample is from the NYU Value-Added Galaxy Catalog (Andreas A. Berlind et al 2006). They use the Hashed-Oct-Tree (HOT) code (Warren & Salmon 1993) to make an N-body simulation with the Lambda-Cold Dark Matter (LCDM) cosmological model, with $\Omega_m = 0.3$, $\Omega_\lambda = 0.7$, $\Omega_b = 0.04$, $h = H_0(100km/s/Mpc) = 0.7$, $n = 1.0$, and $\sigma_8 = 0.9$. $\Omega_m$ is the total matter mass. Density is in units of the critical density for closure, $\rho_0 = 3H_0^2/8\pi G$. $\Omega_b$ and $\Omega_\lambda$ are densities of baryons and dark energy at the present day. The Hubble constant $H_0 = 100 km/s/Mpc$, $n$ is the simulation's initial density perturbation spectral index, while $\sigma_8$ is the rms linear mass fluctuation within a sphere of radius 8 Mpc/h extrapolated to z = 0. This model is in agreement with a wide variety of cosmological observations (Blanton et al. 2005). Initial conditions were set up using the transfer function calculated for this cosmological model by CMBFAST (Seljak & Zaldarriaga 1996). Then they used the friends-of-friends (FOF) algorithm to identify galaxy halos in simulation, with FOF length equal to 0.2 times the mean inter-particle separation. After getting haloes, based on the Halo Occupation Distribution (HOD, which is a model to get the probability distribution *P(N/M)* that a halo (dark matter particles cluster) of mass *M* contains *N* galaxies), they created the NYU Value-Added Galaxy Catalog employing some other restrictions, such as relations between spatial and velocity distributions of galaxies and dark matter within halos (Berlind & Weinberg 2002).

The second mock sample is Millennium Run semi-analytic galaxy catalogue (Croton et al 2006) based on the Millennium Run LCDM N-body simulation (Springel et al. 2005). The Millennium Simulation used revised GADGET2 (Croton et al 2006) code and also used the "TreePM" (pure dark matter code, Bagla 2002) method to evaluate gravitational forces. It is a combination of a hierarchical "tree" algorithm and a classical, Fourier transform particle-mesh method. The following cosmological parameters are from Springel's paper (Springel et al 2005): $\Omega_m = \Omega_{dm} + \Omega_b = 0.25$, $\Omega_b = 0.045$, h =0.73, $\Omega_\lambda = 0.75$, n = 1, and $\sigma_8 = 0.9$. Those parameter values are consistent with a combined analysis of the galaxy surveys and first year WMAP (Springel et al 2005) data.

The Catalogues only include galaxies above our magnitude completeness limit ($M_r - 5\log h = -16.6$ and $M_B - 5\log h = -15.6$), for a total of about 9 million galaxies in the full simulation box (500 Mpc/h on a side).

We also created a random sample with the same criteria as the SDSS data, such as volume geometry, spatial density, and selection functions (window functions). The random sample is used for calibrating the MST, and we anticipate the random sample should be very different from the observed sample on most scales, as the observed sample does show some structures (such as filaments), which cannot be found in the random sample.

In our research we use non-equal triangles (faster to calculate) to approximate the surface of components, as in Figure 6.

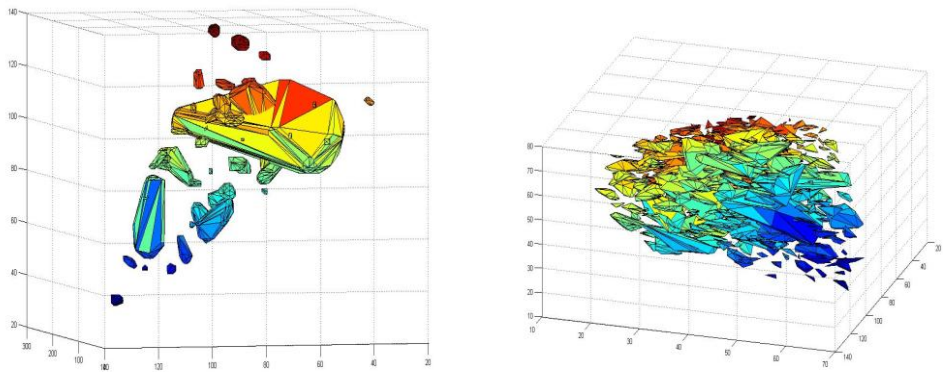

Figure 6. Our surface triangulation application. Two examples of using surface triangulation on SDSS DR5 sample on two different smoothing scales (shown with different orientations).

### 3.4 Standard deviation

To estimate errors of random mock samples, we choose 12 random samples with different seeds (initial conditions) when we calculate the metric distance between the observed sample and the random mock samples. We also extract 12 NYU samples from the same cubic simulation but with different orientations (and minimized overlapping (~20% overlap) of the sample regions) to get deviation of the NYU sample. For the MPA sample and observed sample we cannot make subsamples (due to the limited size of the original data) and thus they have no errorbars (we borrow the errorbars from the NYU sample for some figures).

### 4. Results

We chose 8~30 Mpc as the range of smoothing lengths (FWHM) and analyzed the clumps with 10 threshold values equal spaced from maximum to minim value of the map. From Equation (4) we get the overall filament value (each clump has same weight regardless of the different size). To illustrate the filamentation property of the observed data, we compare observed image with uniform image (f=0, in other words, no filamentation at all). Figure 7 shows the calculated filament values for the observational SDSS data.

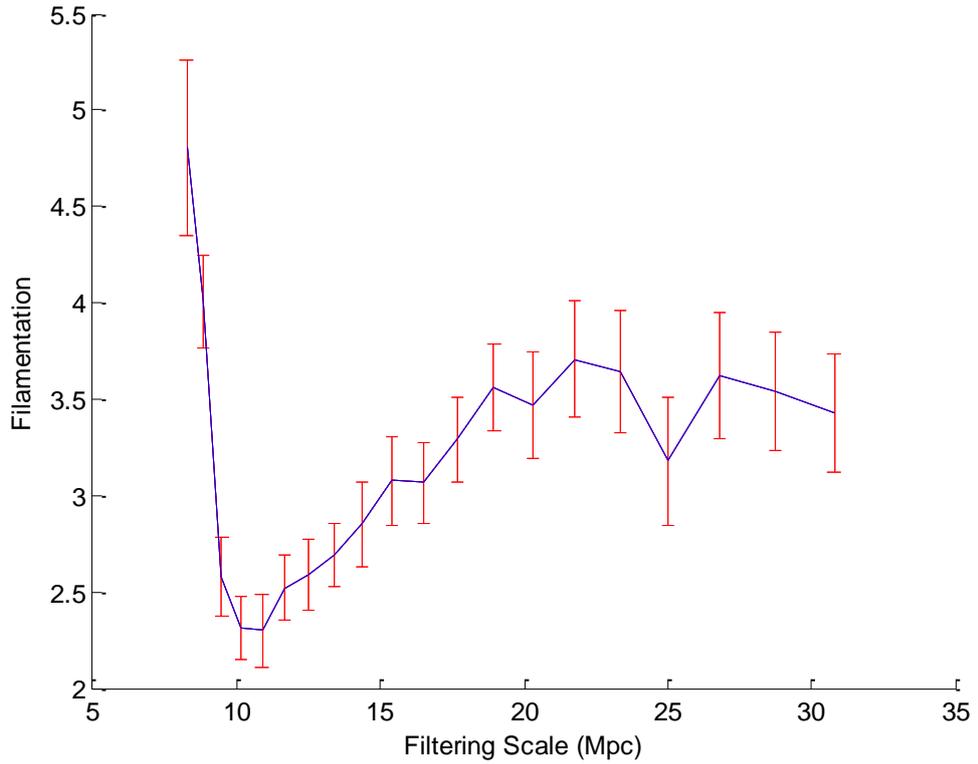

Fig 7. Results of the observed sample compared with uniform image on different filtering scales, using the filamentation measure. The x-axis value is smoothing length, which ranges from 8 Mpc to 31 Mpc. Error bars are "borrowed" from NYU mock sample results.

We can see there is a turning point around 10 Mpc scale. With the definition of filamentation index, clumps at first become less filamentary (from 5.3 to 2.4) with the increasing smoothing scale, but after 10 Mpc smoothing scale they become more filamentary (from 2.4 to 3.5). This suggests the possible existence of large filaments in the SDSS sample. Then function is flat (around 3.5) at 20 Mpc scale and larger.

Now we look at the difference among mock samples and the observed sample. First we compare all samples with the observed sample (Eq. 4).

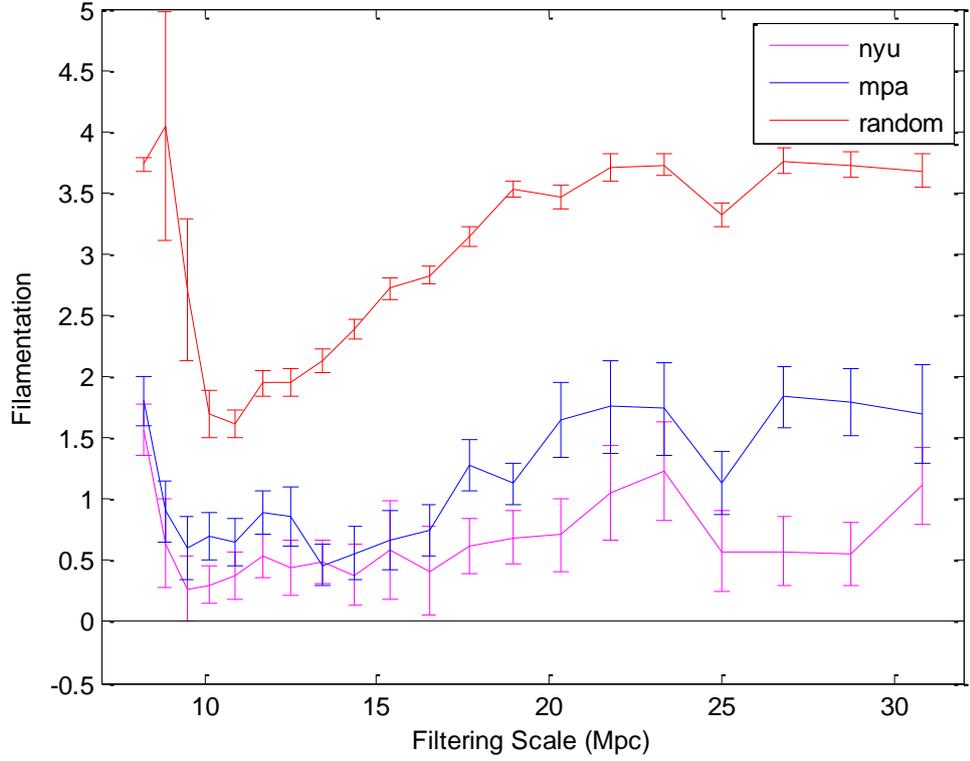

Figure 8. Metric distance for filament function. We also plot the straight thick solid line representing the zero value of distance from the observed sample.

We can distinguish filament value of random sample from other samples very well ($\geq 6\sigma$ difference) and find that the NYU sample behaves slightly better than the MPA sample (around $2\sigma$).

We now know the metric distance between the mock and observed samples (shown in the y-axis of Figure 8, calculated from Equation (4)), but we do not know if mock samples have greater or less of filamentation than the observed sample. We only know the "distance", with no sign. So we set *p*=1 in Equation (4), then we will get a new metric distance, with sign. The results are shown in Figure 9.

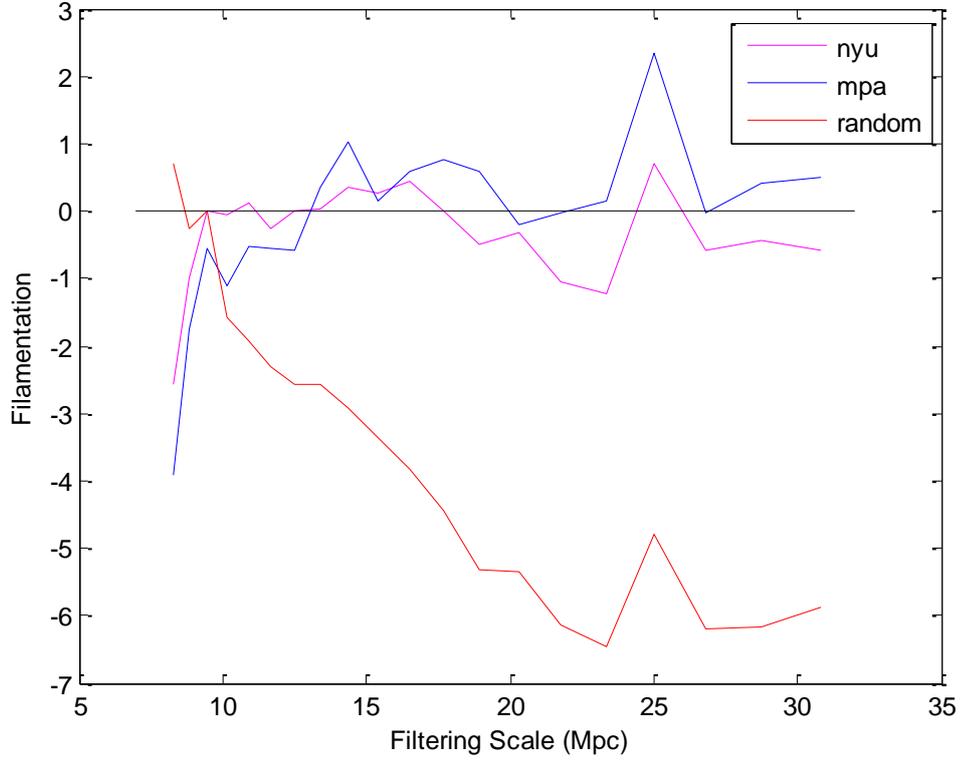

Figure 9. New metric distances with *p*=1 in Equation (4).

This new information shows that NYU tends to have less filamentation, while MPA generally has more than the observed sample, filament function reflects that NYU is closer to observed sample than MPA samples (more than $3\sigma$ difference for filament function). In the small scale (<10 Mpc), the filament value of both mock samples are smaller (negative metric distance) than the observed sample, interestingly random sample get a larger filament value than observed sample on small scales.

## 5. Conclusions

We have used our filament index definition on multiple scales to study the filamentation of galaxy distributions. The technique gives a detailed filamentation description of galaxy distributions in metric space, on scales from approximately 8 Mpc to 30 Mpc showing statistically strong differences among the samples. We also find that filament function has minimums around 10 Mpc in Figures 8 & 9, reflecting that there are some filament structures above 10 Mpc scale in SDSS galaxy distribution.

The key motivation of this research is to supplement traditional tools with a more informative way of quantifying the similarity in the "visual" filamentation properties between simulations and the observed Universe. It was demonstrated that two N-body simulations have done a good job of approximating our Universe and that NYUr is significantly closer to the observed sample than MPAr. We have the expected result that the random sample is much different from all other samples at virtually all scales for filamentation.

The Millennium Run simulation used in this paper was carried out by the Virgo Supercomputing Consortium at the Computing Center of the Max-Planck Society in Garching. The semi-analytic galaxy catalog is publicly available at http://www.mpa-garching.mpg.de/galform/agnpaper. We thank Andreas A. Berlind for providing the NYU Mock Galaxy Catalog.